\documentclass[11pt]{article}
\usepackage{fullpage}
\usepackage{graphicx} 	% extended graphics package
\usepackage{amsmath} 	% math package
\usepackage{array} 	% extended styles for tables
\usepackage{hhline} 	% extended styles for tables
\usepackage{times}      % times fonts look better
\usepackage{amsfonts} 
\usepackage{verbatim}
\usepackage{url}
\usepackage{pifont}
\usepackage[nocompress]{cite}
\usepackage{color}
\usepackage{xspace}
\usepackage{ifpdf}
\ifpdf
\else
\fi

\newcommand{\urlBiBTeX}[1]{\url{#1}} %%% needed for URL to print nice
\newcommand{\bulletpoint}[1]{\medskip\noindent{\em #1}.} %%% easier to customize

\def\allitems{24,385\xspace}
\def\allfeedback{184,804\xspace}
\def\allsellers{1,239\xspace}
\def\startdate{February 3, 2012\xspace}
\def\enddate{July 24, 2012\xspace}
\def\topsellerfb{4,847\xspace}
\def\finalizeearly{20,884\xspace}
\def\badfe{342\xspace}
\def\customlisting{745\xspace}
\def\coresellers{112\xspace}
\def\website{\url{https://arima.cylab.cmu.edu/sr/}\xspace}

\newcommand{\timestamp}{{\bf Working paper}\\First version: May 4, 2012.\\This version: \today.\\\mbox{\small $\ $Id: paper.tex 1654 2012-11-28 17:04:37Z nicolasc $\ $}}

\begin{document}

\title{\bf Traveling the Silk Road:
A measurement analysis\\ of a large anonymous online marketplace
}

\author{
Nicolas Christin\\
{\em Carnegie Mellon INI/CyLab}\\
\url{nicolasc@cmu.edu}
}
\date{\timestamp}

\maketitle
\begin{abstract}
We perform a comprehensive measurement analysis of Silk Road, an
anonymous, international online marketplace that operates as a Tor
hidden service and uses Bitcoin as its exchange currency. We gather and
analyze data over eight months between the end of 2011 and 2012,
including daily crawls of the marketplace for nearly six months in
2012. We obtain a detailed picture of the type of goods being sold
on Silk Road, and of the revenues made both by sellers and Silk Road
operators. Through examining over 24,400 separate items sold on the
site, we show that Silk Road is overwhelmingly used as a market for
controlled substances and narcotics, and that most items sold are
available for less than three weeks. The majority of sellers disappears
within roughly three months of their arrival, 
but a core of 112 sellers has been present throughout our measurement
interval. We evaluate the total revenue made by all sellers, from
public listings, to slightly over USD~1.2 million per month; this
corresponds to about USD~92,000 per month in commissions for the Silk
Road operators. We further show that the marketplace has been operating
steadily, with daily sales and number of sellers overall increasing over
our measurement interval. We discuss economic and policy implications of
our analysis and results, including ethical considerations for future
research in this area.
\end{abstract}

\newpage
\section{Introduction}
\label{sec:intro}
``More brazen than anything else by light-years'' is how U.S. Senator
Charles Schumer characterized Silk Road \cite{silkroad}, an online
anonymous marketplace. While a bit of a hyperbole, this
sentiment is characteristic of a certain nervousness among political
leaders when it comes to anonymous networks. The relatively recent
development of usable interfaces to anonymous networks, such as the
``Tor browser bundle,'' has indeed made it extremely easy for anybody
to browse the Internet anonymously, regardless of their technical
background. In turn, anonymous online markets have emerged, making it
quite difficult for law enforcement to identify buyers and sellers. As a
result, these anonymous online markets very often specialize in ``black
market'' goods, such as pornography, weapons or narcotics.

Silk Road is one such anonymous online market. It is not the only
one -- others, such as Black Market Reloaded \cite{BMR}, the Armory
\cite{armory}, or the General Store \cite{generalstore} are or have been 
offering
similar services -- but it gained fame after an article posted on
Gawker \cite{silkroad-gawker}, which resulted in it being noticed by
congressional leaders, who demanded prompt action be taken. It is
also reportedly very large, with estimates mentioned in the Silk Road
online forum \cite{silkroadforums} ranging between 30,000 and 150,000 active
customers.
\begin{figure}
\begin{center}
\includegraphics[width=0.9\columnwidth]{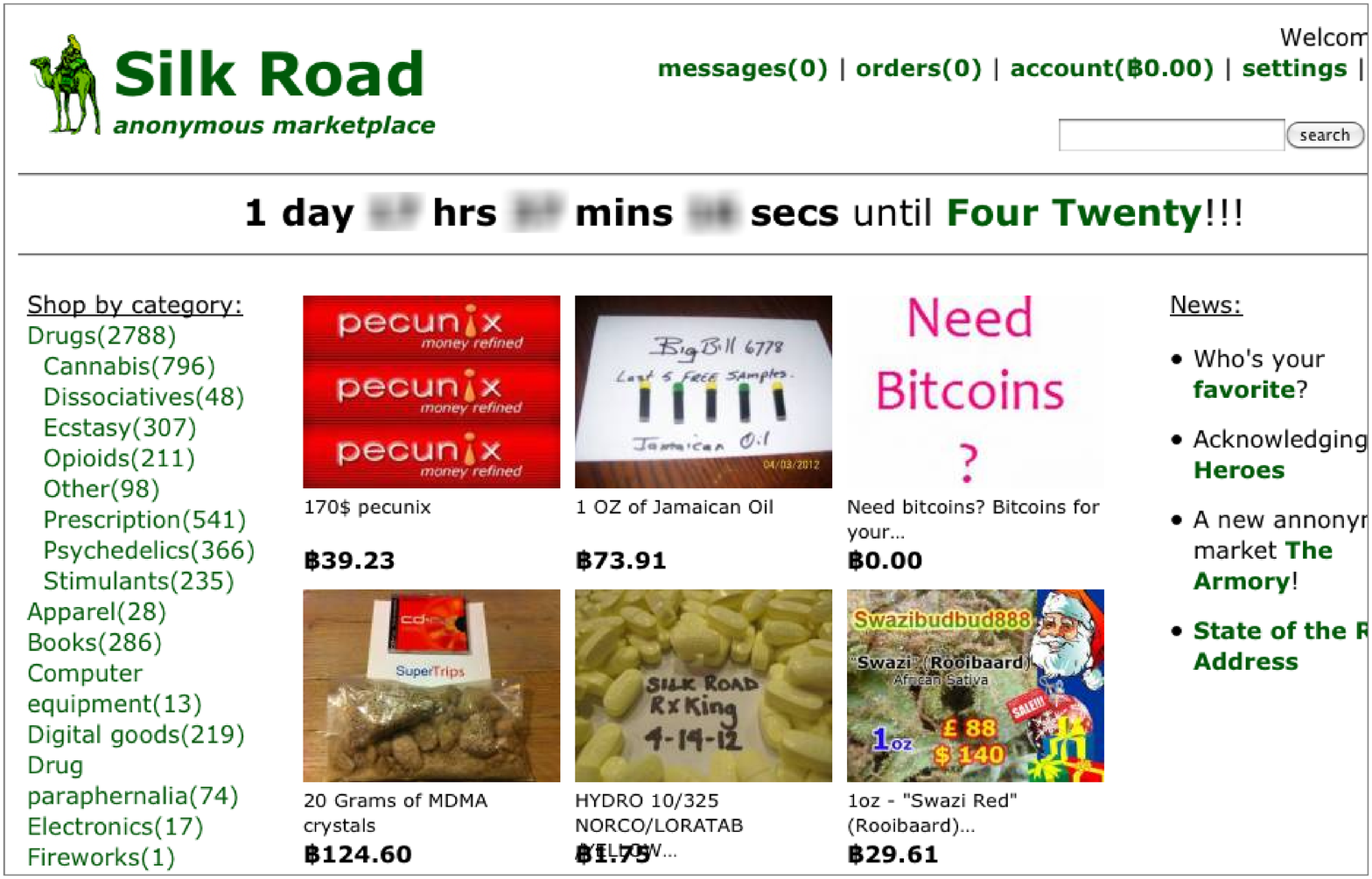}
\end{center}
\caption{\label{fig:silkmain}{\bf Silk Road front page.} The site offers a number of licit and illicit items, with a marked focus on narcotics.}
\end{figure}

Figure~\ref{fig:silkmain} shows the Silk Road front page. The site has
a professional, if minimalist, look, and appears to offer a variety of
goods (e.g., books, digital goods, digital currency...), but seems to
have a clear focus on drugs. Not only do most items listed appear to be
controlled substances, but the screenshot also shows the site advertising a
sale campaign for April 20 -- also known as ``Pot day'' due to the North
American slang for cannabis (four-twenty). 

In this paper, we try to provide a scientific characterization of the
Silk Road marketplace, by gathering a set of controlled measurements
over roughly six months (February 3, 2012 -- July 24, 2012), and
analyzing them. 

Specifically, we offer the following contributions. We
devise a (simple) collection methodology to obtain publicly available
Silk Road market data. We use the data collected to characterize the
items being sold on Silk Road and the seller population. We describe
how items sold and seller population have evolved over time. Using
(mandatory) buyer feedback reports as a proxy for sales, we characterize
sales volumes over our measurement interval. We provide an estimate
of the daily dollar amount of sales conducted on Silk Road, and use
this estimate to infer the amount collected in commission by Silk Road
operators. While we cannot estimate the number of buyers with the
dataset we collect, we show that Silk Road is a relatively significant
market, with a few hundred sellers, and monthly total revenue of about
USD~1.2 million. We also show that Silk Road appears to be growing over
time, albeit not at the exponential rate that is claimed in forums
\cite{silkroadforums}.

The rest of this paper is structured as follows. We start by describing
how Silk Road operates in Section~\ref{sec:background}. We then explain
how we gather our measurements in Section~\ref{sec:collection}. We
report on our measurements analysis in Section~\ref{sec:analysis},
before turning to economic implications in Section~\ref{sec:econ}.
We discuss our findings, reflect on possible 
intervention policies, and ethical considerations 
in Section~\ref{sec:discuss}, outline related work in
Section~\ref{sec:related}, and conclude in Section~\ref{sec:concl}.

\section{Silk Road overview}
\label{sec:background}
Silk Road is an online anonymous marketplace that started its
operations in February 2011 \cite{silkroadforums}. Silk Road is
not, itself, a shop. Instead, it provides infrastructure for sellers
and buyers to conduct transactions in an online environment. In this
respect, Silk Road is more similar to Craigslist, eBay or the Amazon
Marketplace than to Amazon.com. The major difference between Silk
Road and these other marketplaces is that Silk Road focuses on
ensuring, as much as possible, anonymity of both sellers and buyers. In
this section, we summarize the major features of Silk Road through a
description of the steps involved in a typical transaction: accessing
Silk Road, making a purchase, and getting the goods delivered.

\bulletpoint{Accessing Silk Road}
Suppose that Bob ($B$), a prospective buyer, wants to access the
Silk Road marketplace ($SR$). Bob will first need to install a Tor
client on his machine, or use a web proxy to the Tor network (e.g.
\url{http://tor2web.org}) as Silk Road runs only as a Tor hidden
service \cite{Dingledine:USENIX04}. That is, instead of having a DNS
name mapping to a known IP address, Silk Road uses a URL based on
the pseudo-top level domain {\tt .onion}, that can only be resolved
through Tor. At a high level, when Bob's client attempts to contact the
Silk Road server URL (\url{http://silkroadvb5piz3r.onion} at the time
of this writing), Tor nodes set up a rendez-vous point inside the Tor
network so that the client and server can communicate with each other
while maintaining their IP addresses unknown from observers and from each other.

Once connected to the Silk Road website, Bob will need to create an
account. The process is simple and merely involves registering a user
name, password, withdrawal PIN, and answering a CAPTCHA. After this
registration, Bob is presented with the Silk Road front page (see
Figure~\ref{fig:silkmain}) from where he can access all of Silk
Road's public listings. 

\bulletpoint{Public and stealth listings}
Silk Road places relatively few restrictions on the types of goods
sellers can offer. From the Silk Road sellers' guide \cite{silkroad},
\begin{quote}
``Do not list anything who's (sic) purpose is to harm or defraud, such as stolen
items or info, stolen credit cards, counterfeit currency, personal info,
assassinations, and weapons of any kind. Do not list anything related to
pedophilia.''
\end{quote}
Conspicuously absent from the list of prohibited items are prescription
drugs and narcotics, as well as adult pornography and fake identification
documents (e.g., counterfeit driver's licenses). Weapons and ammunition
used to be allowed until March 4, 2012, when they were transferred to 
a sister site called The Armory \cite{armory}, which operated with an
infrastructure similar to that of Silk Road. Interestingly, the Armory
closed in August 2012 reportedly due to a lack of business
\cite{silkroadforums}.

Not all of the Silk Road listings are public. Silk Road supports {\em
stealth listings}, which are not linked from the rest of Silk Road,
and are thus only accessible by buyers who have been given their URL.
Stealth listings are frequently used for {\em custom listings} directed
at specific customers, and established through out-of-band mechanisms
(e.g., private messaging between seller and buyer). Sellers may further
operate in {\em stealth mode}, meaning that their seller page and {\em
all} the pages of the items they have for sale are not linked from other
Silk Road pages. While Silk Road is open to anybody, stealth mode allows
sellers with an established customer base to operate their business as
invitation-only.

\bulletpoint{Making a purchase}
After having perused the items available for sale on Silk Road,
Bob decides to make a purchase from Sarah ($S$), a seller. While Tor ensures
communication anonymity, Silk Road needs to also preserve payment
anonymity. To that effect, Silk Road only supports Bitcoin (BTC,
\cite{bitcoin}) as a trading currency. Bitcoin is a peer-to-peer,
distributed payment system that offers 
its participants to engage in verifiable transactions 
without the need for a central third-party. 
Bob thus needs to first procure Bitcoins, which he
can do from the many online exchanges such as Mt.Gox \cite{mtgox}. Once Bob has Bitcoins, and decides to purchase the item from Sarah, instead of paying Sarah
directly, Bob places the corresponding amount in {\em escrow} with 
Silk Road. Effectively, $B$ pays $SR$, not 
$S$. The escrow mechanism allows the market operator to accurately
compute their commission fees, and to resolve disputes between sellers
and buyers. Silk Road mandates all sellers and buyers use the escrow
system. Failure to do so is punishable by expulsion from the marketplace \cite{silkroad}. 

\bulletpoint{Finalizing} Once the purchase has been made, Sarah must
ship it to Bob. Thus, Sarah needs a physical address where to send the
item. To preserve anonymity, Silk Road recommends to use delivery
addresses that are distinct from the buyer's residence. For instance,
Bob could have the item delivered at Patsy's house, or to a post-office box. 
Once Sarah has
marked the item as shipped, Bob's delivery address is erased from all
records. Once the item reaches its destination, Bob {\em finalizes} the
purchase, that is, he tells Silk Road to release the funds held in escrow to Sarah (i.e., $SR$ now pays $S$), and
leaves feedback about Sarah. Finalizing is mandatory: if Bob forgets
to do so, Silk Road will automatically finalize pending orders after a
set amount of time.

Sellers with more than 35 successful transactions and who
have been active for over a month are allowed to ask their buyers to
{\em finalize early}; that is, to release payment and leave feedback
before they actually receive the item. Due to the potential for abuse, 
Silk Road discourages finalizing
early in general, and prohibits it for new sellers.

Finally, Silk Road enhances transaction anonymity by providing
``tumbler'' services that consist of inserting several dummy, single-use
intermediaries between a payer and a payee. That is, instead of having
a payment correspond to a simple 
transaction chain $B \rightarrow SR \rightarrow S$, the
payment goes through a longer chain $B \rightarrow I_1 \rightarrow
\ldots \rightarrow I_n \rightarrow S$ where $(I_1, \ldots I_n)$ are
one-time-use intermediaries.

\section{Collection methodology}
\label{sec:collection}
We next turn to describing how we collected measurements of the Silk
Road marketplace. We first briefly explain our crawling mechanism,
before outlining some of the challenges we faced with data collection.
We then discuss in detail the data that we gathered.

\subsection{Crawling mechanism}
We registered an account on Silk Road in November 2011, and started
with a few test crawls. We immediately noticed that Silk Road relies
on authentication cookies that can be reused for up to a week without
having to re-authenticate through the login prompt of the website.
Provided we can manually refresh the authentication cookie at least once
per week, this allows us to bypass the CAPTCHA mechanism and automate
our crawls.

We conducted a near-comprehensive crawl of the site on November 29,
2011,\footnote{All dates and times are expressed in Universal Time
Coordinates (UTC).} using HTTrack \cite{httrack}. Specifically, we
crawled all ``item,'' ``user'' (i.e., seller) and ``category'' webpages.
The complete crawl completed in about 48~hours and corresponded to
approximately 244~MB of data, including 124~MB of images. 

Starting on February 3, 2012, and until July 24, 2012, we attempted to
perform daily crawls of the website. We noticed that early in 2012, Silk
Road had moved to inlining images as {\tt base64} tags in each webpage. This
considerably slowed down crawls. Using an incremental mode, that is,
ignoring pages that had not changed from one crawl to the next, each
of these crawls ran, on average, for about 14~hours. The fastest crawl
completed in slightly over 3~hours; the slowest took almost 30~hours,
which resulted in the following daily crawl to be canceled. To avoid
confusion between the time a crawl started, and the time a specific page
was visited, we recorded separate timestamps upon each visit to a given page.

\subsection{Challenges}
Kanich et al. \cite{Kanich:LEET08} emphasize the importance of ensuring
that the target of a measurement experiment is not aware of the
measurement being conducted. Otherwise, the measurement target could
modify their behavior, which would taint the measurements. We thus
waited for a few days after the November crawl to see if the full crawl
had been noticed. Perusing the Silk Road forums \cite{silkroadforums},
we found no mention of the operators noticing us; our account was still
valid and no one contacted us to inquire about our browsing activities.
We concluded that we either had not been detected, or that the operators
did not view our activities as threatening.

We spent some additional effort making our measurements as difficult
to detect as possible. Since all Silk Road traffic is anonymized over
Tor, there is no risk that our IP address could be blacklisted. However,
an identical Tor circuit (on our side) could be repeatedly used if
our crawler keeps the same socket open; this in turn could reveal
our activities if the Silk Road operators monitor the list of Tor
circuits they are running, and realize that a fixed Tor rendez-vous
point is constantly being used. We addressed this potential issue by
ensuring that all circuits, including active circuits, are periodically
discarded and new circuits are built. To further (slightly) obfuscate
our activities, instead of always starting at the same time, we started
each crawl at a random time between 10pm and 1am UTC.

Despite all of these precautions, we had to discard some of our data.
On March 7, 2012 a number of changes were implemented to Silk Road to prevent profiling of the site \cite{silkroadforums}. Whether this was due to Silk Road 
operators noticing our crawls or to other activity is unclear. URL
structure changed: item and users, instead of being referenced by a
linearly increasing numeric identifier, became unique hashes. Fortunately,
these hashes initially simply 
consisted of a substring of the MD5 hash of the numeric identifier,
making it easy to map them to the original identifiers.\footnote{New
identifiers subsequently created are salted hashes with a non-trivial
salt; but those do not map to items and users that had already been
registered on the site when the switch occurred. Thus, we do not need
to find the pre-images of these hashes and can instead simply treat
them as unique identifiers.} More problematically, feedback data, which is
crucial to estimating the volume of sales became aggregated and feedback
timestamps disappeared. That is, instead of having, for an item $G$ sold
by $S$ a list of $n$~feedback messages corresponding to $n$~purchases
of $G$ along with the associated timestamps, Silk Road switched to
presenting a list of 20 feedback messages, undated, across {\em all}
the items sold by $S$. In other words, feedback data became completely
useless. Thankfully, due to very strong pushback from buyers who argued
that per-item feedback was necessary to have confidence in purchases
\cite{silkroadforums}, Silk Road operators reverted to timestamped,
per-item feedback on March 12, 2012. Nevertheless, we had to discard all
feedback data collected between March 7, 2012 and March 12, 2012.

Finally, in several instances, Silk Road went down for maintenance, or
authentication was unsuccessful (e.g., because we had not refreshed
the authentication cookie in time), leading to a few sporadic days
of missing data. The largest gaps are two eight-day gaps between April
10, 2012 and April 17, 2012 due to an accidental suspension of the
collection infrastructure; and between July 12, 2012 and July 19,2012,
due to an accidental deletion of the authentication cookie.

\subsection{Data collected}
\begin{figure}
\begin{center}
\includegraphics[width=0.98\columnwidth]{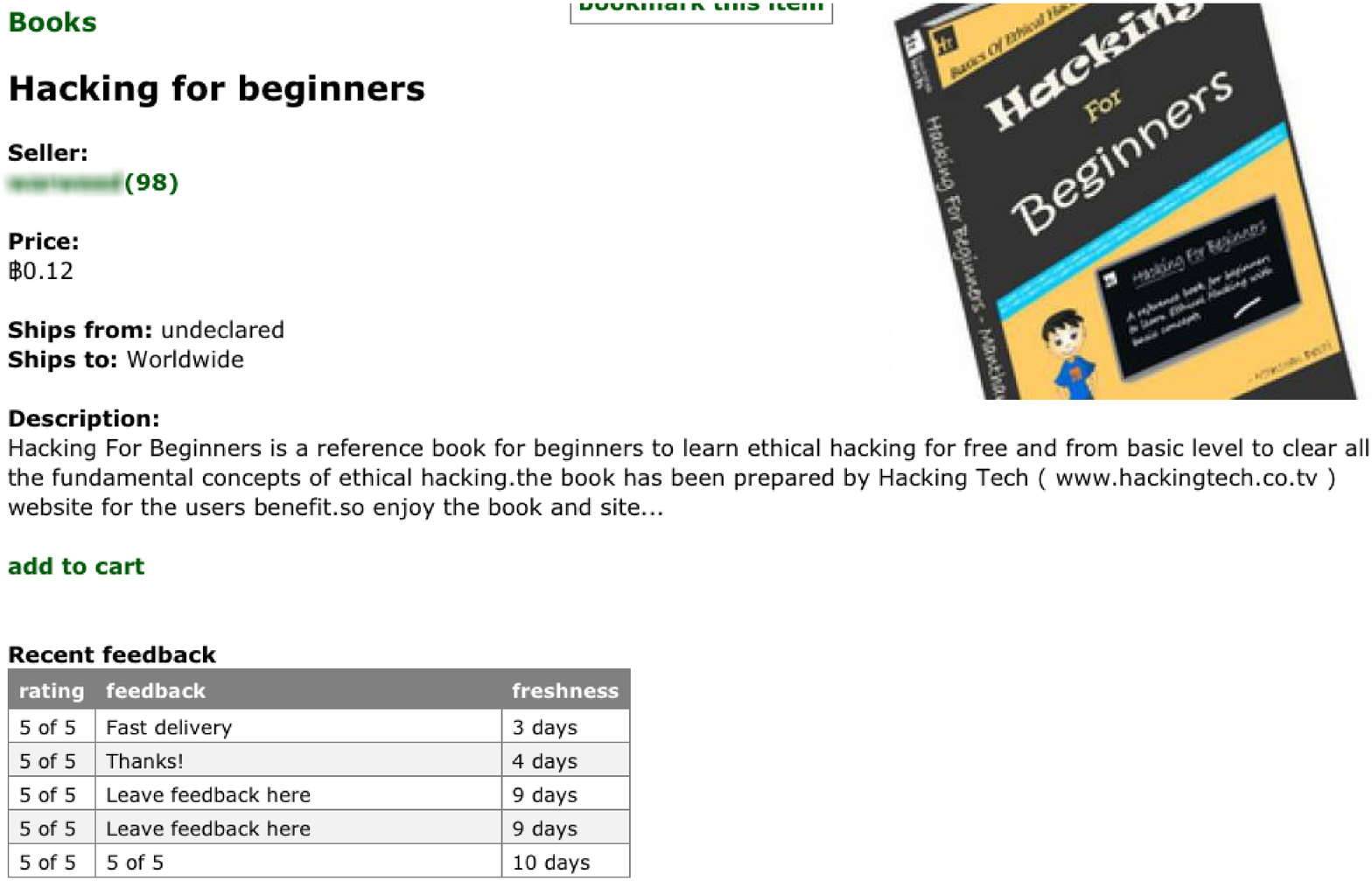}
\end{center}
\caption{\label{fig:item}{\bf Silk Road item page.} Each item page contains seller, price, and shipping information, as well as buyer feedback on the item.}
\end{figure}
We only collect data that is both publicly accessible over the Tor
network, and linked from other Silk Road pages. 
That is, we do not collect buyer data, as buyers do not have
public ``buyer pages.'' We also do not collect stealth listings, or data
about sellers when they operate in stealth mode.

We primarily focus data collection on ``item pages,'' that is, pages
describing the goods being sold on Silk Road. We show an example
in Figure~\ref{fig:item}. Each item page is bound to a unique item
identifier as part of its URL (integer until March 7, 2012, 10-digit
hash afterwards), and contains the name of the item (``Hacking for
beginners'' in Figure~\ref{fig:item}), a picture, the category in
which the item fits (e.g., ``Books''), seller information (a name,
percentage of positive feedback, and a hyperlink denoting the seller
unique identifier), price (e.g., 0.12~BTC), shipping information, item
description, and buyer feedback. We gather all of this information for
each item we crawl, and record a timestamp (in UNIX epoch time) every
time the page is visited.

\bulletpoint{Feedback data} 
Each piece of feedback consists of three fields: a rating between 1 and
5, a textual description of the feedback, and the age of the feedback.
Feedback age is expressed in minutes, hours, days or months,
depending on how old the feedback is. Hence, we can timestamp much
more accurately feedback recently given at the time of the crawl, than
older feedback. This is one of the reasons for crawling Silk Road
daily: the age of feedback less than a day old can be quite precisely
pinpointed. 

We record feedback in two different manners. For each crawl of Silk
Road started at time~$t$ and lasting until $t+\tau$ ($\tau > 0$) ,
we record all feedback present on the site in a separate database
$\mathcal{D}_t$, thereby getting a {\em snapshot} of the feedback
amassed until time~$t+\tau$. This method may miss some feedback. For
instance, if we crawl an item page at time $t+\tau_1$, and a customer
leaves feedback at time $t+\tau_2$ with $\tau_1 < \tau_2 < \tau$,
that customer's feedback will not be recorded as part of the time-$t$
snapshot. Furthermore, timestamps of feedback given long before $t$ may
be very approximate.

To address this issue, we also record, in a database $\mathcal{D}$,
novel feedback from one crawl to the next, that is, feedback for which
text did not previously appear in our records for this specific item.
This method guarantees that feedback timestamps are as accurate as
possible (since they are recorded as soon as the feedback is observed).
Furthermore, we can capture nearly all the feedback present on the
site, without worrying about collection gaps. A drawback of this method
is that it may overestimate the amount of feedback when there are
feedback updates. In particular, new buyers are sometimes asked to
finalize early, that is, to send feedback immediately after the online
transaction is completed and before receiving goods. They may elect to
update the feedback after delivery of the goods purchased, which can
be weeks later. When this happens, the original feedback is replaced
on the website by the new feedback, and the timestamp is updated.
However, $\mathcal{D}$ contains both the original, and the updated
feedback(s), even though only one sale occurred. 

Maintaining both a family $(\mathcal{D}_t)$ of database of snapshots
of the site, and a cumulative database $\mathcal{D}$ allows us to have
lower and upper bounds on the amount of feedback posted on the site,
which in turn is a useful indicator of sales.

\section{Marketplace characteristics}
\label{sec:analysis}
We next provide an overview of the types of goods being sold in Silk
Road, before discussing seller characteristics.

\subsection{What is being sold?} 
Items offered on Silk Road are grouped by categories. There are
approximately 220~distinct categories, ranging from digital goods 
to various kinds of narcotics or prescription
medicine.
\begin{figure}
\includegraphics[width=0.49\columnwidth]{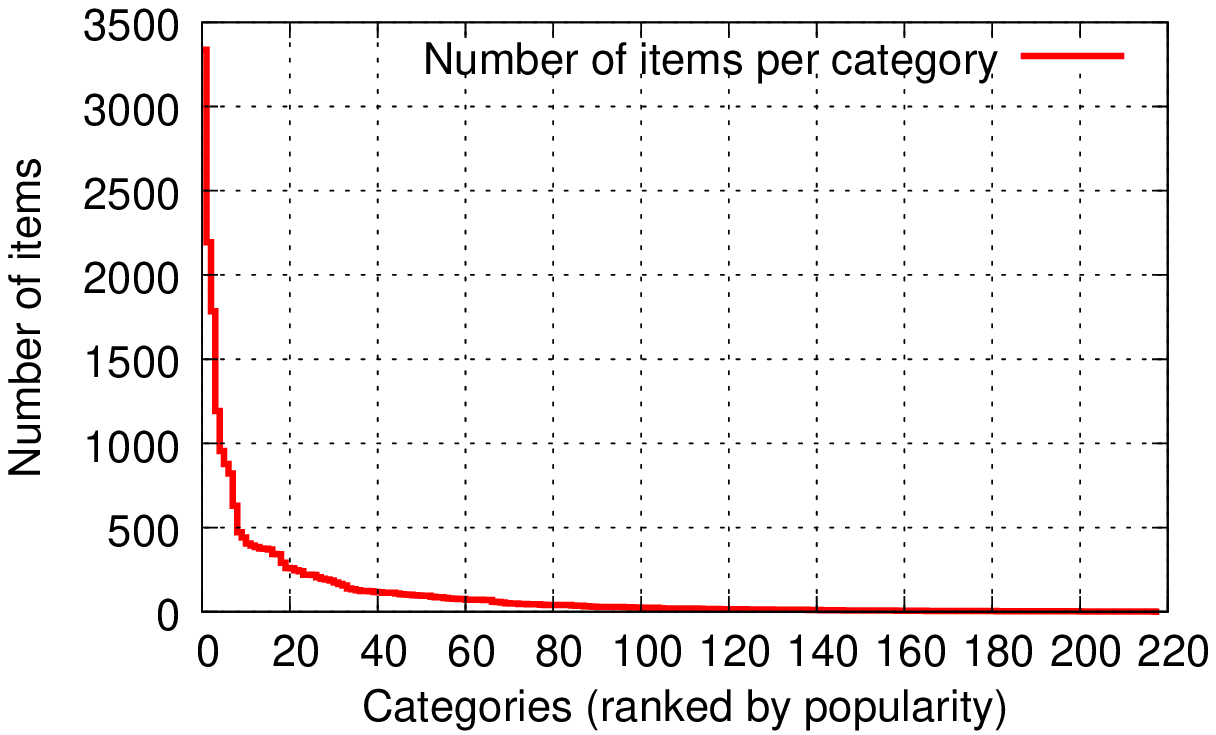}
\includegraphics[width=0.49\columnwidth]{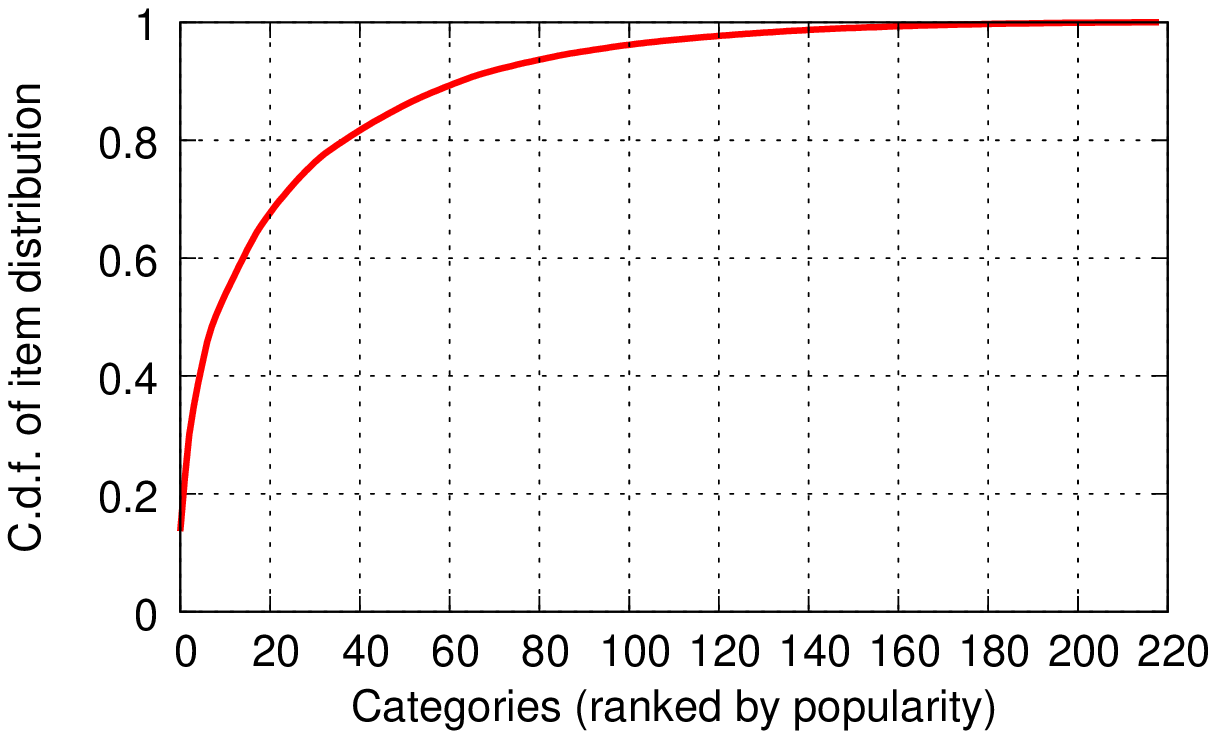}
\caption{\label{fig:category} {\bf Distribution of items per category.} The plots show the number of items in each category, ordered by decreasing popularity (left) and the cumulative distribution of all items over all categories (right). The 20 most popular categories represent over 2/3 of all items available.}
\end{figure}
In Figure~\ref{fig:category}, we plot, on the left-hand side, for each
category, the number of items sold in that category, over the data
collected from \startdate through \enddate. For readability we
ordered categories by decreasing popularity. In total, we found \allitems
unique items being sold over that period. While a few
categories seem to hold the most items, Silk Road, like other online
marketplaces, exhibits a long-tail behavior, where a large number of items appear to be unique. This is confirmed
by the right-hand graph, where we plot the cumulative distribution
of items as a function of the number of categories considered. The
right-hand graph shows that over two thirds of all products sold on Silk
Road during our data collection interval belong to one of the top 20
categories, but that, after that, the cumulative fraction of items only slowly converges to 1.

\begin{table}[t]
\begin{center}
\begin{tabular}{lrr}\hline
{\bf Category} & {\bf \#. items} & {\bf Pct.}\\\hline
Weed    &3338    &13.7\%\\
Drugs   &2194    &9.0\%\\
Prescription    &1784    &7.3\%\\
Benzos  &1193    &4.9\%\\
Books   &955     &3.9\%\\
Cannabis        &877     &3.6\%\\
Hash    &820     &3.4\%\\
Cocaine &630     &2.6\%\\
Pills   &473     &1.9\%\\
Blotter (LSD) &440     &1.8\%\\
Money   &405     &1.7\%\\
MDMA (ecstasy)&393     &1.6\%\\
Erotica &385     &1.6\%\\
Steroids, PEDs  &376     &1.5\%\\
Seeds   &374     &1.5\%\\
Heroin  &370     &1.5\%\\
DMT     &343     &1.4\%\\
Opioids &342     &1.4\%\\
Stimulants      &291     &1.2\%\\
Digital goods   &260     &1.1\%\\\hline
\end{tabular}
\caption{\label{tab:category} {\bf Top 20 categories in terms of items available.} Products sold on Silk Road are mostly listed as narcotics or controlled substances.}
\end{center}
\end{table}
In Table~\ref{tab:category}, we take a closer look at the top~20
categories per number of item offered. ``Weed'' (i.e., marijuana) is the
most popular item on Silk Road, followed by ``Drugs,'' which encompass
any sort of narcotics or prescription medicine the seller did not want
further classified. Prescription drugs, and ``Benzos,'' colloquial
term for benzodiazepines, which include prescription medicines like
Valium and other drugs used for insomnia and anxiety treatment, are
also highly popular. The four most popular categories are all linked to
drugs; nine of the top ten, and sixteen out of the top twenty are
drug-related. In other words, Silk Road is mostly a drug store, even
though it also caters some other products. Finally, among narcotics,
even though such a classification is somewhat arbitrary, Silk Road
appears to have more inventory in ``soft drugs'' (e.g., weed, cannabis,
hash, seeds) than ``hard drugs'' (e.g., opiates); this presumably simply
reflects market demand.

\bulletpoint{Item availability} 
\begin{figure}
\begin{center}
\includegraphics[width=0.75\columnwidth]{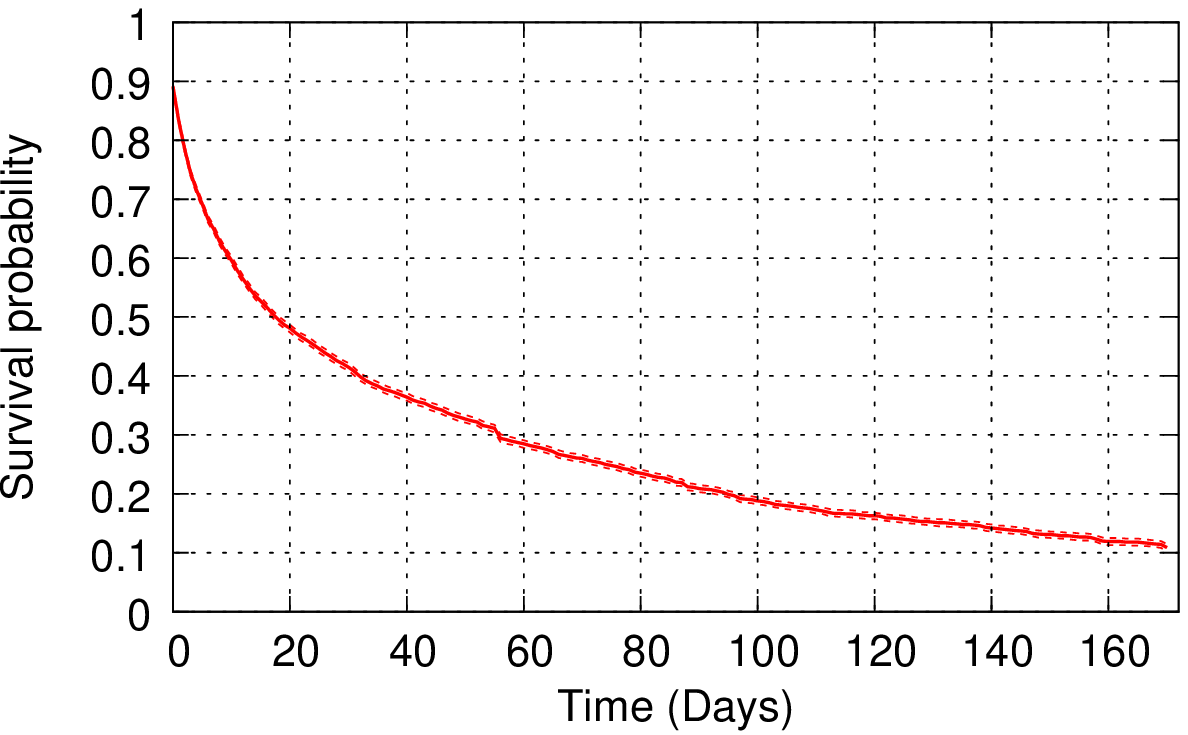}
\caption{\label{fig:item_churn_survival} {\bf Probability that a given
item will be available on the site as a function of time (in days).} The
dotted lines correspond to the 95\% confidence interval. Most items are
only available for limited periods of time, with a majority of items
being available for less than three weeks.}
\end{center}
\end{figure}
We estimate item availability by first recording the first time we saw
an item being listed, and the last time we saw it listed. Items may
have been listed and de-listed several times in the meantime; here we
are only looking at the overall lifespan of an item, regardless of
its transient availability. To account for the large number of
items still available when we stop collection on \enddate, we use a
standard Kaplan-Meier estimator \cite{KM58} to compute the ``survival
probability,'' $S_{\mbox{item}}(\tau)$, of a given item after $\tau$~days, that is,
the probability an item is listed for more than $\tau$ days.
$S_{\mbox{item}}(\tau)$ is plotted in Figure~\ref{fig:item_churn_survival} along with
95\% confidence intervals.

We discover that a majority of items disappear within three weeks of
their first being listed ($S_{\mbox{item}}(21)=0.473$, 95\% c.i. $(0.466, 0.479)$);
and more than 25\% of items disappear within three days 
($S_{\mbox{item}}(3)=0.745$, 95\% c.i. $(0.739, 0.750)$). On the other hand, there
are also a few very long-lived items (on the right-hand side of the
graph) that have been present for the entire collection interval. There
may be two different explanations for the relatively short lifespan
of each item: vendors may run out of stock quickly and de-list their
items, possibly re-listing them later under a slightly different name
resulting in a different item page, or they may elect to make them
stealth listings as soon as they have established a customer base.

\bulletpoint{Custom listings} Finally, public custom listings are
relatively rare. Out of the \allitems items we observed, only \customlisting 
were
explicitly marked as ``custom listings.'' This is undoubtedly a lower
bound, as custom listings should be stealth listings, except when sellers are running tests.

\subsection{Who is selling?}
Due to the anonymous nature of Silk Road, it is impossible to discern
whether certain sellers use multiple seller pages; economically, this is not an attractive proposition, as there is a fee associated with opening a seller account. Likewise, several
sellers in the physical world may offer their goods through a unique
seller page on Silk Road, although this would certainly be a clear
indication of a business partnership. In this discussion we will equate
``sellers'' with distinct seller pages.
\begin{figure}
\begin{center}
\includegraphics[width=0.75\columnwidth]{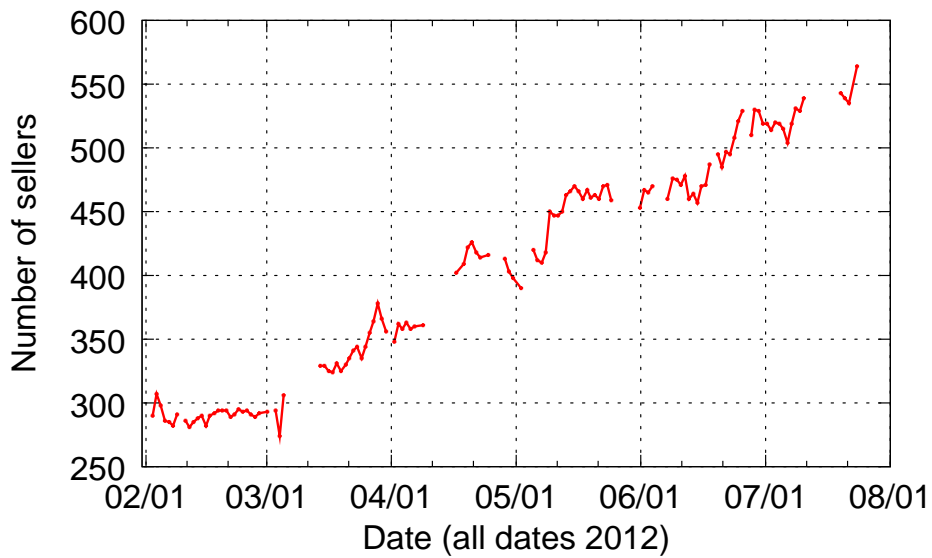}
\caption{\label{fig:seller_count} {\bf Evolution of the number of sellers.} As a point of comparison, there were only 220 sellers on Nov. 29, 2011.}
\end{center}
\end{figure}

\bulletpoint{Evolution over time}
In Figure~\ref{fig:seller_count}, we plot the evolution of the number of
sellers on Silk Road over time, between \startdate
until our last daily crawl (\enddate). 

On \enddate, we found 564 distinct sellers
with at least one item listed for sale on Silk Road.
This is a marked increased compared to the 220 sellers 
we had observed during our initial crawl 
on November 29, 2011 (not shown on the figure). The gaps 
in the figure correspond to data collection gaps. 
 An interesting spike occurs
around April 20, 2012. April 20 featured a large promotional sale on
Silk Road to mark ``Pot Day.'' It appears that a number of sellers
entered the marketplace in the week or two prior to this operation; and a non-negligible number left immediately afterwards.
Also, the Silk Road forums indicate that one of the top sellers went
on hiatus on March 12, 2012. It is unclear whether this played a role
in the marked increase of the number of sellers since that time -- i.e.,
whether newcomers attempted to fill the void. The 
number of active sellers has been continuously
increasing, at least since early March. A linear regression fit
in Figure~\ref{fig:seller_count} gives $y=1.674x+265.605$ ($R^2=0.969$), where $x$ is in
days, $y$ the number of active sellers, and the time origin is set to the beginning of our measurement interval.
In other words, the increase in the number of sellers appears linear,
with about 50~new active sellers each month.

\begin{figure}
\begin{center}
\includegraphics[width=0.75\columnwidth]{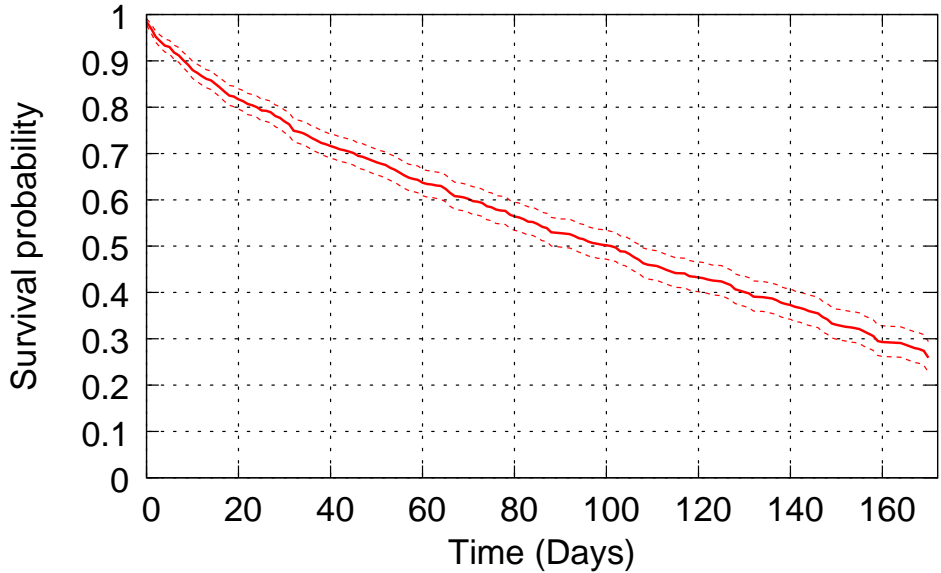}
\caption{\label{fig:seller_churn_survival} {\bf Probability a seller
will continue publicly selling items on the site as a function of time
(in days).} The dotted lines denote the 95\% confidence interval.
About half of the sellers leave the site within 100~days of initial
appearance; around a fifth of the sellers stay for less than three
weeks.}
\end{center}
\end{figure}

However, this simple regression analysis does not accurately reflect
the fact that many sellers actually leave the marketplace. Over the
measurement interval \startdate -- \enddate, we indeed found
\allsellers distinct sellers were at one point or another publicly listed.
We use again a Kaplan-Meier estimator to plot the probability
$S_{\mbox{seller}}(\tau)$ that a given seller will remain active
on the site for at least $\tau$ days. Similar to our discussion
about item availability, we consider a seller is active by
subtracting the first time they were seen from the last time
they were seen, regardless of their coming and going in between.
Figure~\ref{fig:seller_churn_survival} plots the function
$S_{\mbox{seller}}$, along with 95\% confidence intervals.
Compared to item turnover, seller turnover is rather modest. A
majority of sellers stay on the site for approximately 100 days
($S_{\mbox{seller}}(101)=0.500$, 95\% c.i. $(0.469,0.532)$); about
a fifth of all sellers are present for less than three weeks
($S_{\mbox{seller}}(21)=0.813$, 95\% c.i. $(0.791,0.836)$). 
A ``core'' of \coresellers~sellers (9\% of all sellers) were present for the
entire measurement interval. All of this data point to relative seller
stability.

\begin{table}
\begin{center}
\begin{tabular}{lrclr}\cline{1-2}\cline{4-5}
\multicolumn{2}{c}{\bf Origin} & & \multicolumn{2}{c}{\bf Acceptable destinations}\\\cline{1-2}\cline{4-5}
{\bf Country} & {\bf Pct.\hfill}&\ &
{\bf Country/Region} & {\bf Pct.}\\\cline{1-2}\cline{4-5}
U.S.A.  & 43.83\%       &\ & Worldwide       &49.67\%\\
Undeclared  & 16.29\%   &\ & U.S.A.     &35.15\%\\
U.K.    &10.15\%        &\ & European Union  &6.19\%\\
Netherlands     &6.52\% &\ & Canada  &6.05\%\\
Canada  &5.89\%         &\ & U.K.    &3.66\%\\
Germany &4.51\%         &\ & Australia&2.87\%\\
Australia&3.19\%        &\ & World. excpt. U.S.A.    &1.39\%\\
India   &1.23\%         &\ & Germany &1.03\%\\
Italy   &1.03\%         &\ & Norway  &0.70\%\\
China   &0.98\%         &\ & Switzerland     &0.62\%\\
Spain   &0.94\%         &\ & New Zealand     &0.56\%\\
France  &0.82\%         &\ & Undeclared      &0.26\%
\\\cline{1-2}\cline{4-5}
\end{tabular}
\caption{\label{tab:shipments} {\bf Top 12 most frequent shipping origins (left), and acceptable shipping destinations (right).} Certain sellers ship to multiple destinations, hence totals may exceed 100\%.}
\end{center}
\end{table}

\bulletpoint{Geographic location}
We next inspect the
advertised origins and acceptable shipping destinations
for each of the \allitems items. Table~\ref{tab:shipments} shows the top 12 locations for
both origin and destinations. Some items ship to multiple destinations
(e.g., Norway, Switzerland and European Union) so that the total of the
rightmost column adds up to more than 100\%. Most items ship from the
United States, with the United Kingdom a distant second. The Netherlands
are also strongly represented, which is not necessarily surprising given
the relatively permissive narcotics laws in the country. We note a clear
bias toward English-speaking countries which represent almost two-thirds
of all listed origins. This is not surprising since all
communications on Silk Road are in English.

More surprisingly, we note that a majority of items ship worldwide, in
spite of the nature of the items, as discussed above. One would think
that sellers may be reluctant to ship narcotics across borders. It
turns out not to be the case, for a couple of reasons. First, sellers
with an established reputation may often demand that buyers pay upon
purchase, and before delivery of the item. If the item is not delivered,
the buyer may have very little recourse, particularly if they have not
established a strong reputation in the marketplace. While, as discussed
in Section~\ref{sec:background}, Silk Road offers an escrow service,
disputes arising after ``early finalization'' are considerably harder to
mediate by the marketplace operators. Second, the quantities being
sold are generally rather small (e.g., a few grams of marijuana),
and tracing the senders may be a very difficult task as they can
use couriers to mail the items rather than going to the post-office themselves. Third, most sellers use
techniques to make package inspection unlikely -- e.g., vacuum sealing,
``professional-looking'' envelopes with typed destination addresses
\cite{silkroad}. In other words, sellers can expand their customer base
at a relatively low risk for them. Economic incentives justify worldwide
shipping, especially since sellers can factor in their selling price the
risk of package seizure, and accordingly offer loyal customers at least
partial refund guarantees.

\bulletpoint{Seller ranking} 
As discussed above, we observed \allsellers unique seller accounts on Silk Road
between \startdate and \enddate, with between 281 and 564
active at a time.
\begin{figure}
\begin{center}
\includegraphics[width=0.75\columnwidth]{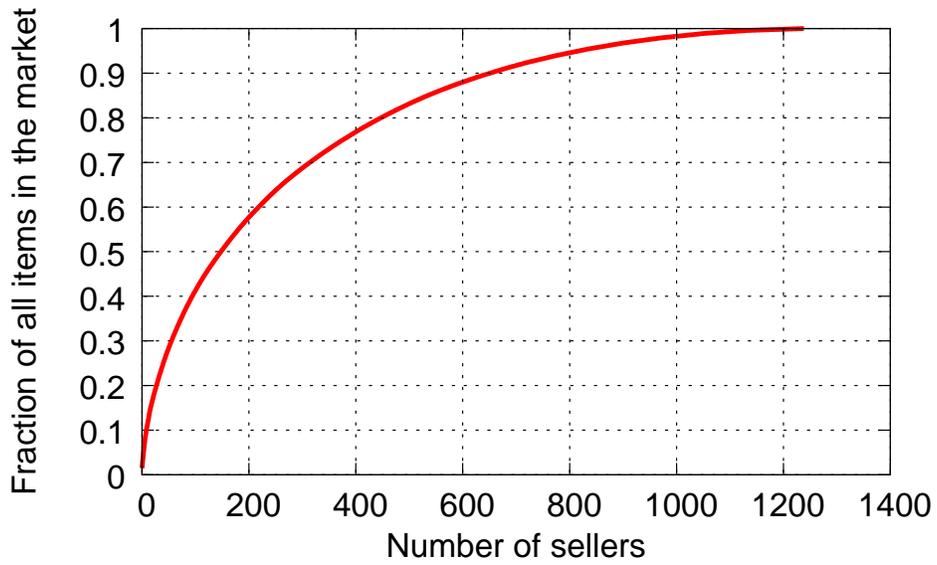}
\caption{\label{fig:sellers_by_item} {\bf Proportion of items in the marketplace as a function of the number of sellers.} We observe fairly high diversity, with each seller selling at most 1.5\% of the total number of items in Silk Road.}
\end{center}
\end{figure}
Figure~\ref{fig:sellers_by_item} shows the proportion of items in the
marketplace as a function of the number of sellers. 
No seller is selling a large inventory. 
Instead, each seller accounts, at
most, for 1.5\% of the total number of items found on Silk Road.

\begin{figure}
\includegraphics[width=0.49\columnwidth]{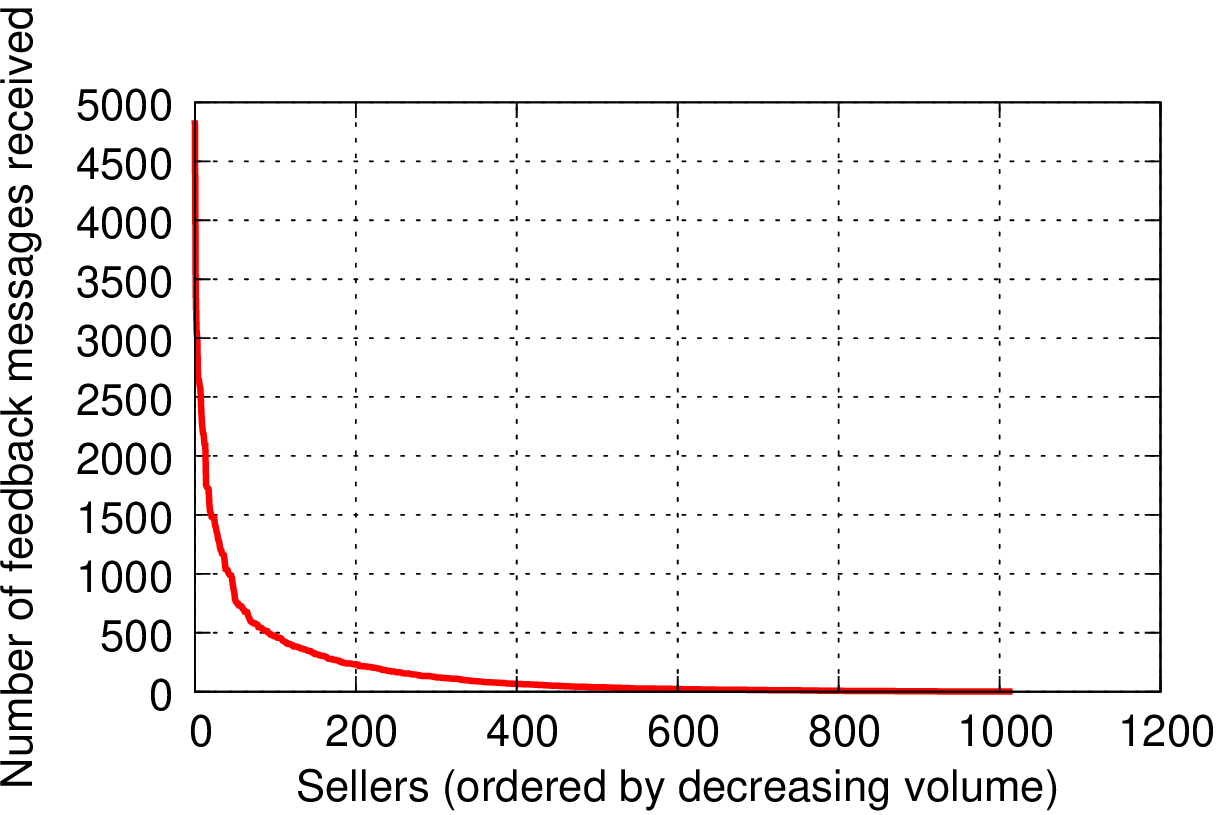}
\includegraphics[width=0.49\columnwidth]{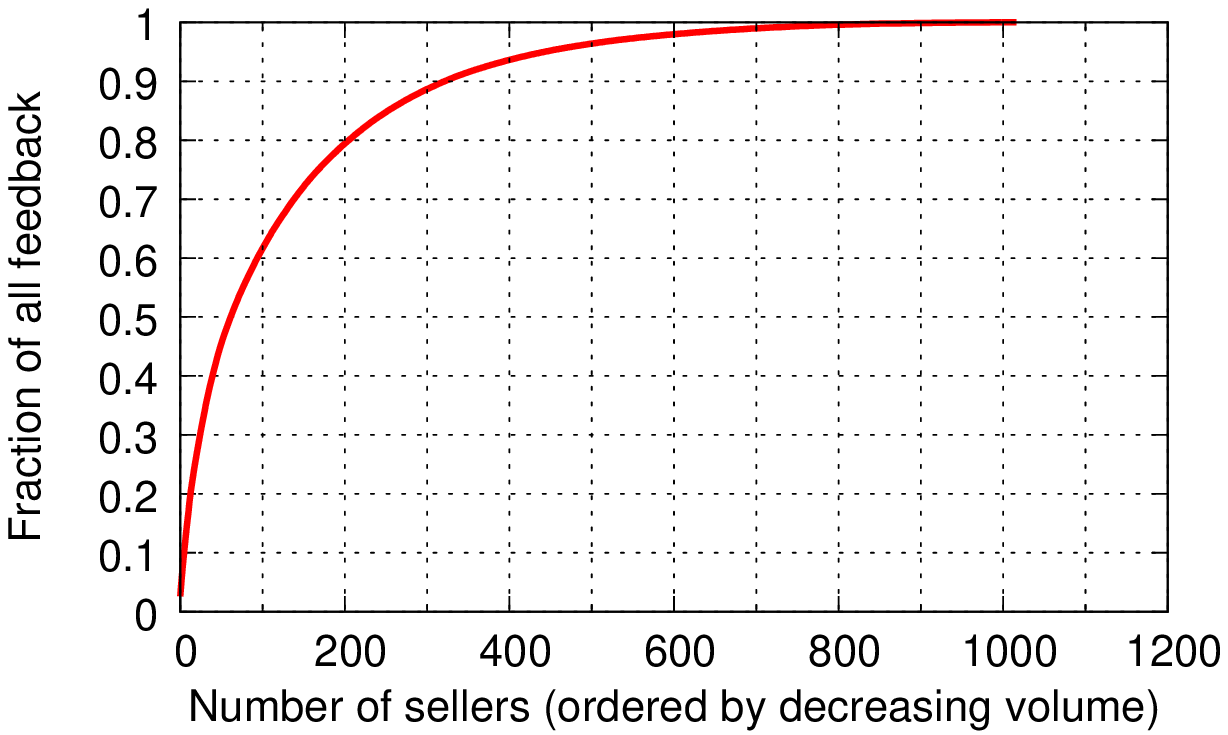}
\caption{\label{fig:seller_vol} {\bf Number of feedback received per seller (left) and cumulative fraction of all feedback observed over the collection interval as a function of the number of sellers (right).} The top 100 sellers are responsible for approximately 60\% of all feedback gathered.}
\end{figure}
While it does not appear like a given seller (or group of sellers) sells
a significant proportion of items overall, it could be the
case that a few selected items sell in large quantities. As discussed in
Section~\ref{sec:collection} we use the amount of feedback collected as
a proxy for the number of sales made. In Figure~\ref{fig:seller_vol}, we
plot, on the left-hand side, the number of feedback received per seller
(where sellers are ranked in decreasing amount of feedback received),
and on the right-hand side the cumulative fraction of total feedback as
a function of the number of sellers considered. These plots show that
a few sellers indeed receive a large amount of feedback, in absolute
terms. 
For instance, the seller with the largest volume of sales received \topsellerfb
feedback messages over the six months we monitored. 
However, the market is quite spread out between sellers, as can be seen
on the right-hand graph. Roughly 100~sellers correspond to 60\%
of all feedback gathered. 

\begin{figure}
\begin{center}
\includegraphics[width=0.75\columnwidth]{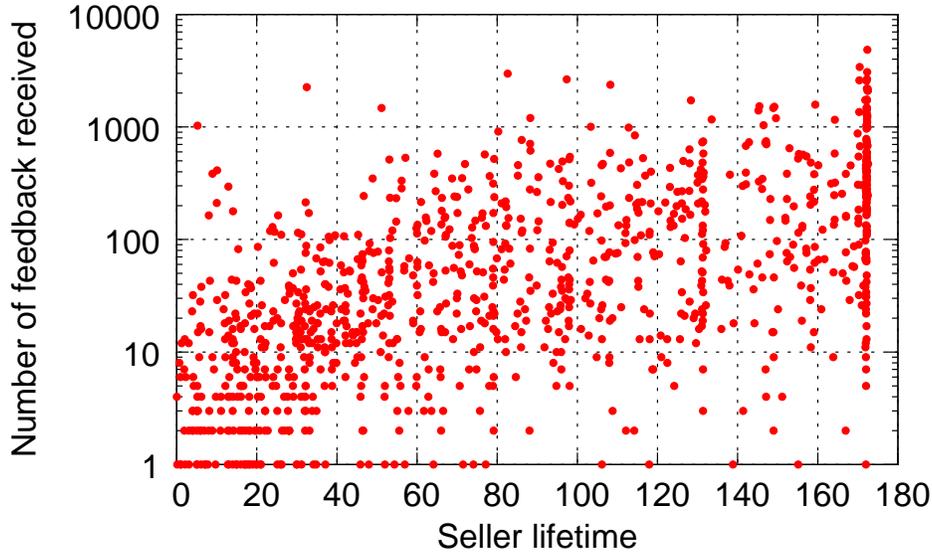}
\caption{\label{fig:lifetime_vs_feedback} {\bf Amount of feedback received by sellers vs. the amount of time sellers have been listed on the site.} There is a modest positive correlation between the two quantities with Pearson's correlation coefficient $r\approx 0.39$. 
}
\end{center}
\end{figure}

Considering we previously observed that
\coresellers~sellers have been present throughout our measurements,
we hypothesized that this core was perhaps the most active. In
Figure~\ref{fig:lifetime_vs_feedback}, we use a scatter-plot to graph
the amount of feedback received (in logarithmic scale) by each seller
against the number of days the seller has been present on the site.
While the graph suggests a positive correlation between the two
quantities, that correlation appears quite weak. This is confirmed by
computing Pearson's correlation coefficient between the two datasets: we
get $r\approx 0.39$. In other words, there are a number of sellers that
do not appear to stay very long on the site, but amass a large number of
transactions, while on the other hand, some of the ``old timers'' appear
to relatively rarely engage in transactions. These results may be due to
the sellers electing to go in stealth mode after having built up a large
enough user base, or to list most of their products as stealth listings.

\subsection{Customer satisfaction}
\begin{table}
\begin{center}
\begin{tabular}{lrr}
{\bf Rating} & {\bf Number} & {\bf Pct.}\\\hline
5/5 &178341& 96.5\%\\
4/5 &2442 & 1.3\%\\
3/5 &1447 & 0.8\%\\
2/5 &520 & 0.3\%\\
1/5 &2053 & 1.1\%\\\hline 
\end{tabular}
\caption{\label{tab:feedback} {\bf Distribution of feedback ratings.} A vast majority of transactions seems to proceed to the satisfaction of the buyers.}
\end{center}
\end{table}
We next discuss customer satisfaction. On a site like Silk Road, where
seller anonymity is guaranteed, and no legal recourse exists
against scammers,
 one would expect
a certain amount of deception. Most transactions seem, however, to generate excellent
feedback from buyers. Table~\ref{tab:feedback} provides a breakdown
of the feedback ratings from \allfeedback feedback instances we collected.
97.8\% of feedback posted was positive (4 or 5 on a scale of 1 to 5). In
contrast, only 1.4\% of feedback was negative (1 or 2 on the same scale).

Thus, it appears at first glance that Silk Road sellers are highly
reliable; or, at the very least, that the escrow system used is effective at policing the marketplace. We caution, however, against too rapid an interpretation of this
result. First, a study by Resnick and Zeckhauser \cite{Resnick:AAM2002}
shows that Internet users in general disproportionately use positive
feedback when rating online experiences. In fact, over 99\% of the
feedback in the eBay corpus used in their study \cite{Resnick:AAM2002} 
was positive. Second,
not all transactions have feedback reported. Indeed, a number of
transactions are made ``out of escrow,'' i.e., directly between a
seller and a buyer. For those, there is no feedback mechanism, nor any
oversight possible from the Silk Road operators. We suspect most of the
scams\footnote{The Silk Road operators in fact warn that buyers relying
on out of escrow transactions ``have been scammed \cite{silkroad}.''}
occur ``out of escrow,'' and of course, no feedback for these
transactions is reported since they technically do not exist (from the
market operator standpoint). 

\bulletpoint{Finalizing early} 
We observe that \finalizeearly instances of feedback contain variations of
``F.E.,'' or ``finalizing early,'' accounting for spelling variations (``finalize'' vs. ``finalise'') and word order (``early finalization'' vs ``finalize early''). 
This shows that
finalizing early is a rather common practice on Silk Road. There does
not appear to be significantly more problems reported with feedback
including such strings (only \badfe of them map to a rating of 1 or 2).
This seems to show that established sellers that are offered the option
of requesting early finalization from their customers do not abuse that
privilege.

\section{Economic aspects}
\label{sec:econ}
We next present a brief discussion of economic indicators on Silk Road. We start by looking at the evolution of the prices of a basket of items. We then turn to estimating transaction volumes that occur on Silk Road. 
\subsection{Inflation}
All transactions on Silk Road are using Bitcoins (BTC). Bitcoin has been a
notoriously volatile currency, going from 1~BTC being worth around
30 cents in January 2011 to 1~BTC reaching USD~31.90 in an intra-day
high on June 8, 2011, and declining rapidly back to approximately 
1~BTC$\approx$USD~2.5 in late October 2011. 

\begin{figure}[tbhp]
\begin{center}
\includegraphics[width=0.75\columnwidth]{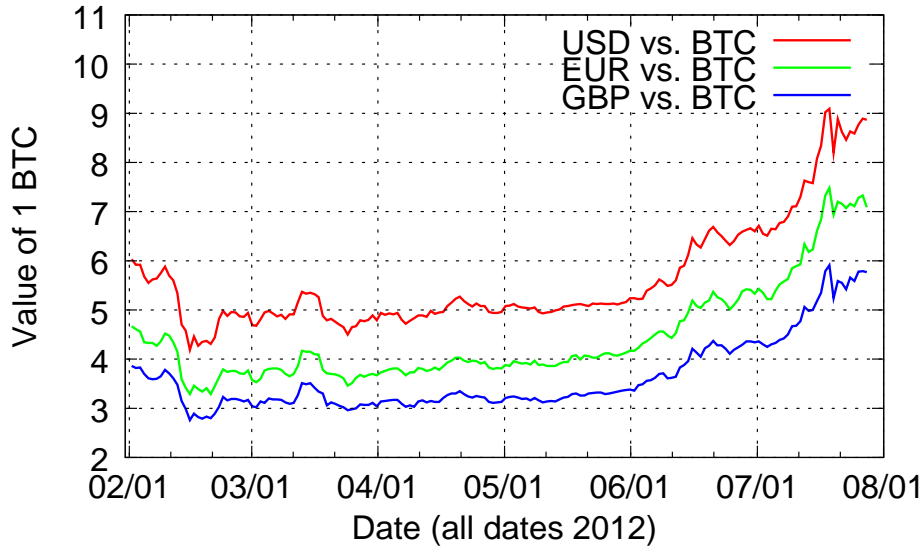}
\caption{\label{fig:bitcoin} {\bf Evolution of the value of a Bitcoin in the three major currencies used in the countries of the sellers operating on Silk Road, over our collection interval.} Each point corresponds to a weighted average over a given day. Data is from Mt.Gox \cite{mtgox}.}
\end{center}
\end{figure}
In Figure~\ref{fig:bitcoin}, we plot the evolution of the exchange rate
of the Bitcoin against the three major currencies that sellers use
in their countries, over the duration of our measurements. As can be
seen in the figure, the Bitcoin exchange rate has remained relatively
stable between the end of February and early May, oscillating around
1~BTC$\approx$USD~5, and corresponding values in euros and pounds. Since
then, the Bitcoin has notably appreciated, reaching close to USD~9 since
mid-July 2012, with relatively large fluctuations.
\begin{figure}[bthp]
\begin{center}
\includegraphics[width=0.49\columnwidth]{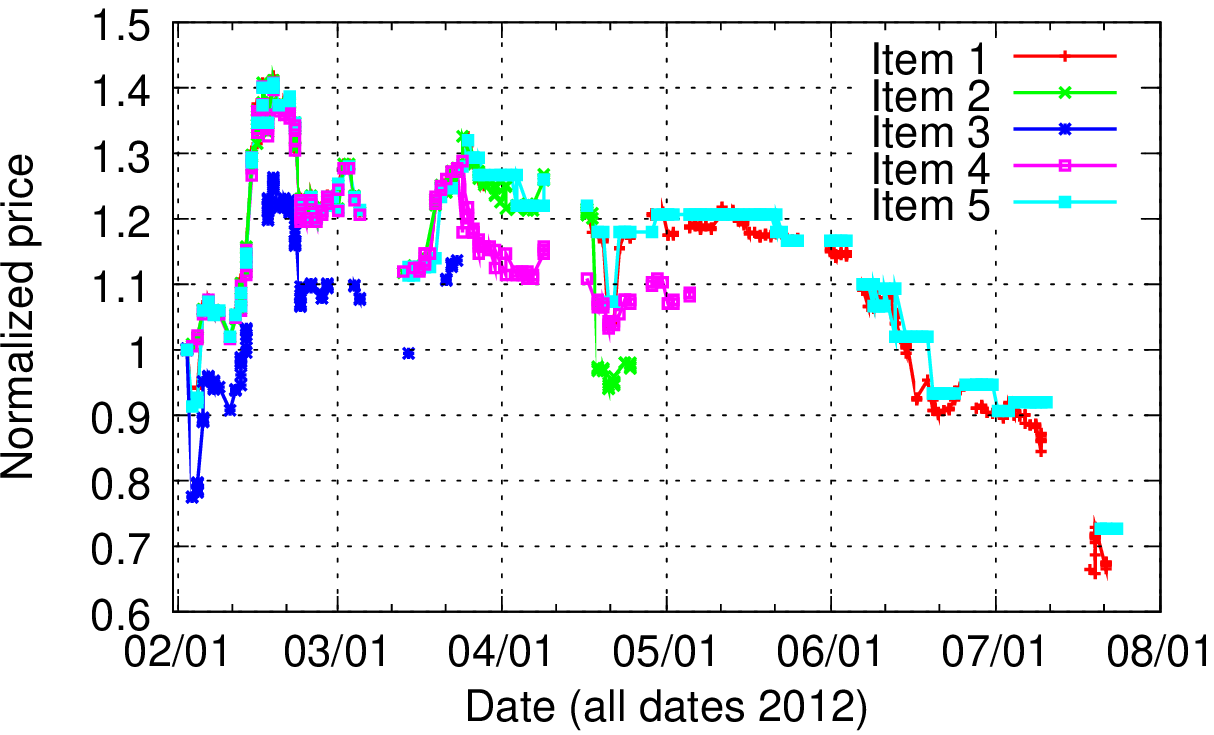}
\includegraphics[width=0.49\columnwidth]{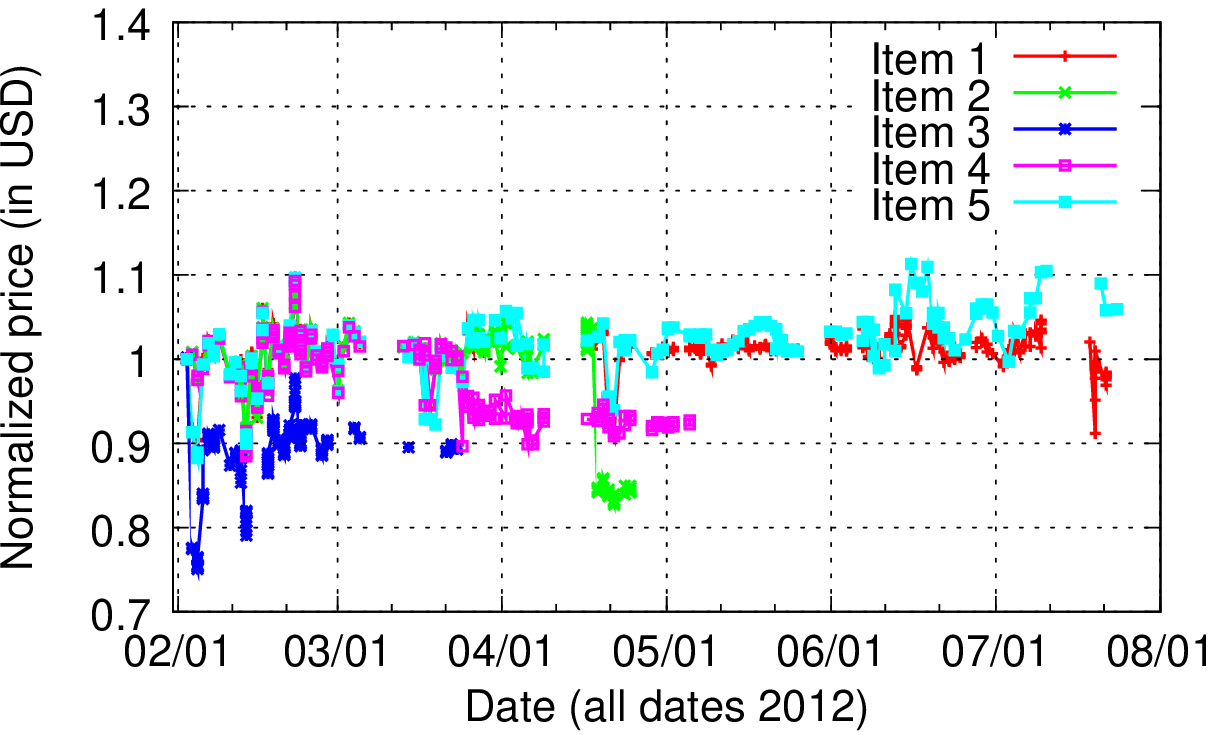}
\caption{\label{fig:basket} {\bf Evolution of the normalized price of the five most sold items on Silk Road in BTC (left) and US dollar (right).} The evolution closely mirrors the evolution of the Bitcoin exchange rates, suggesting little inflation for these items over the time interval considered.}
\end{center}
\end{figure}

We look at the evolution of the normalized price of the five
best-selling items on Silk Road, corresponding to between 1,025 and
1,590 pieces of feedback collected. These five items are all narcotics, but correspond to four 
different categories, and have prices ranging from 1~BTC to approximately 
50~BTC. For any time~$t$, the normalized price compared to origin $t_0$
is defined as $P(t)/P(t_0)$.

We plot the normalized prices of these five items in
Figure~\ref{fig:basket}. On the left, we observe that they quite closely
mirror the fluctuations of the Bitcoin exchange rate: as the Bitcoin
appreciates, the prices drop; conversely, a drop in the Bitcoin value
results in a price increase. 

Silk Road provides automatic pegging of
prices to the US dollar if sellers so desire. 
On the right, we plot the normalized price in US dollars. 
The plot confirms that there is generally no evidence of rapid inflation (or
deflation) on Silk Road as all prices remain roughly within $\pm$10\% of
their original price.
We notice
a drop in price prior to April 20 for all items. Indeed, this price
decrease appears to be part of a promotional campaign for ``Pot
Day.''
A second observation is that item~2 stops being
sold immediately after April 20. The last time it is observed on the
site is April 25, before being de-listed. From discussions in Silk Road
forums \cite{silkroadforums}, it appears that the seller of that item
abruptly left the marketplace, potentially leaving a large number of
paid, finalized early, orders unfulfilled. In other words, there is
suspicion of a ``whitewashing attack \cite{Feldman:JSAC06},'' whereby
a seller creates an excellent reputation, before using that reputation
to defraud users and leave the system. In hindsight, the 20\% drop in
price occurring just prior to April 20 was considerably steeper
than all the other promotional discounts. This could have been an indicator
that the seller was not intending on fulfilling their orders and was
instead artificially lowering prices in hopes of attracting large numbers
of customers to defraud.

\subsection{Transaction volumes}
We next provide an estimate of the total amount of daily sales realized
in Silk Road. Obtaining this estimate is problematic, because, as
explained in Section~\ref{sec:collection}, feedback data is relatively
noisy. Feedback can be updated, which results in a timestamp update, and
in discarding the old feedback; and feedback is not always issued
at the time the item is purchased, but instead, when it is received.

We decided to use averages over a sliding window of size 29 days to try
to assess the daily sales volumes. The averaging mechanism smoothes
out potential issues due to delayed feedback. The reason why we limit
ourselves to 29~days is that timestamps on feedback older than 30~days
(i.e., one month) are extremely approximate, with potential errors of 29
days in the worst case, while timestamps on feedback less than a month
old are accurate within 24 hours.

\begin{figure}[tbhp]
\begin{center}
\includegraphics[width=0.75\columnwidth]{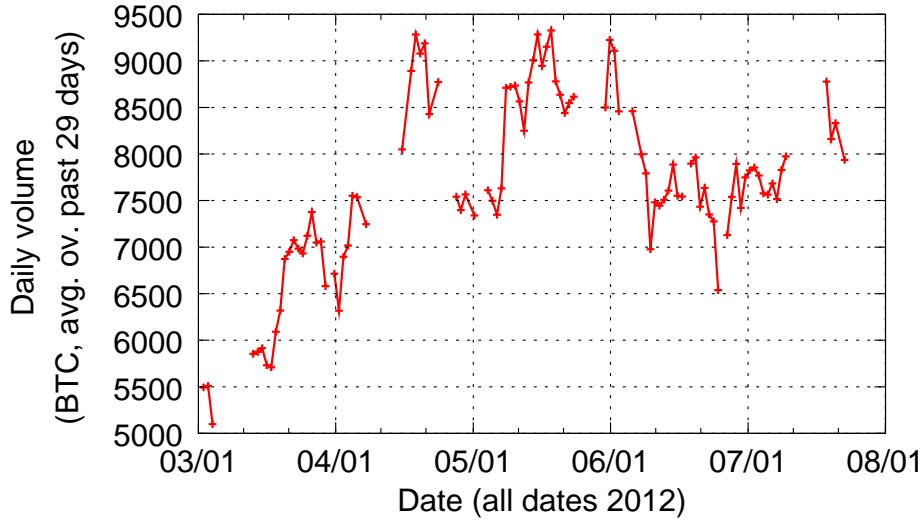}
\caption{\label{fig:daily_sales} {\bf Estimate of the total amount of daily sales (in BTC) occurring on Silk Road.} Each point corresponds to an average over the prior 29~days.}
\end{center}
\end{figure}

We present our results in Figure~\ref{fig:daily_sales}. Each point
corresponds to the average daily sales volumes over the past 29 days.\footnote{We use a 29-day as opposed to a 30-day window, as 30-day old feedback may be marked as ``30 days old'' or as ``one month old'' depending on the exact time when it is collected, resulting in significant errors in feedback counts.}
The sales volumes are computed by, for each item, counting the number
of feedback gathered about the item over the past 29 days, and
multiplying this number by the average price of the item over the last
29 days. The estimated daily sales volumes are then simply taken as
the resulting sales volume divided by 29. 

We observe that the total volume of sales has been increasing quite
significantly, going from approximately 6,000 BTC/day to approximately
9,500 BTC/day, before seemingly retreating down to 7,000
BTC/day.

The latter decrease is, however, an artifact of the Bitcoin sharply
appreciating against all major currencies, rather than an indication
of a drop in sales. Over our entire collection interval, the daily
volume of sales approaches 7,665 BTC/day. 

Converting to US dollars, using the Bitcoin exchange rate plotted in
Figure~\ref{fig:bitcoin}, we obtain a total sales volume of over USD
1.22~million per month when averaged over our measurement interval.
This would correspond to an annual revenue of
close to 15~million USD for the entire marketplace.

\subsection{Operator commissions}
Silk Road operators collect a commission on all sales realized on the
website. The commission schedule was originally set at 6.23\%
of the sales price. In January 2012, a tiered commission schedule was
established, using a model reminiscent of eBay's fee structure. 
\begin{table}
\begin{center}
\begin{tabular}{lr}
{\bf Item price} & {\bf Op. commission}\\\hline
first \$50 & 10\%\\
\$50.01 - \$150 & 8.5\%\\ 
\$150.01 - \$300 & 6\%\\ 
\$300.01 - \$500 & 3\%\\ 
\$500.01 - \$1000 & 2\%\\
over \$1000 & 1.5\% \\\hline
\end{tabular}
\caption{\label{tab:commission} {\bf Silk Road operator commission schedule (from January 2012).} Prior to the establishment of this schedule, the commission was at a flat 6.23\% rate.}
\end{center}
\end{table}
The schedule, described in Table~\ref{tab:commission}, is
based on amounts in US Dollars, not in Bitcoins. The first USD~50 of a
sale are charged a 10\% commission fee, then, from~USD 50.01 to USD~150,
the fee becomes 8.5\%, and further decreases as the value of the item
sold increases.

\begin{figure}[tbhp]
\begin{center}
\includegraphics[width=0.75\columnwidth]{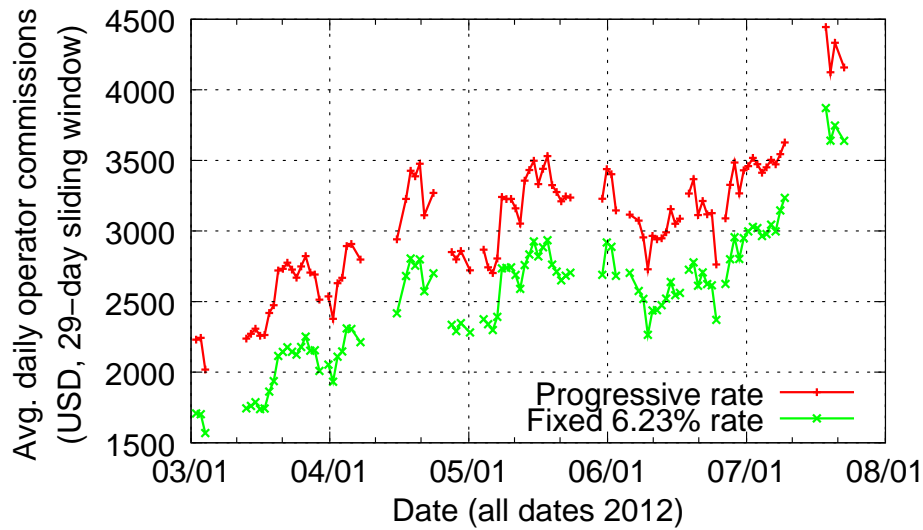}
\caption{\label{fig:daily_commission} {\bf Estimate of the daily commission collected by Silk Road operators (in USD).} Each point corresponds to an average over the prior 29~days. The progressive rate schedule is given in Table~\ref{tab:commission}.}
\end{center}
\end{figure}
In Figure~\ref{fig:daily_commission}, 
we plot an estimate of the daily commissions collected by Silk
Road operators as a function of time. We simply reuse the previous
estimates, and apply both the fixed 6.23\% rate, and the schedule of
Table~\ref{tab:commission} to each item. 
We find that the new schedule turns out to yield on average a commission
corresponding to approximately 7.4\% of the item price.

Using the new schedule we find that Silk Road operators have seen
their commissions increase from over USD~2,200/day in March 2012 to
roughly USD~4,000/day in late July 2012. In other words, even though the
volume in Bitcoins may have decreased due to the Bitcoin rising 
against the US dollar, the transaction volume in US dollars, and
the corresponding commissions, have significantly increased over our
measurement interval. Over the entire measurement interval, we compute
that Silk Road operators collect, an average, roughly 92,000 USD per
month in commissions. Stated differently, projected 
over a year, Silk Road operators'
revenue would probably be around 1.1~million US dollars. 

\subsection{Silk Road in the Bitcoin economy}
Finally, we were interested in computing the share of Silk Road trade in
the overall Bitcoin economy. This is a notoriously difficult quantity
to estimate, as quantifying the total volume of Bitcoin transactions
is, itself, challenging. In particular, the use of tumblers and mixers
implies that each Silk Road transaction corresponds to a large number of
Bitcoin transactions.

As a potential indicator, we suggest to compare the estimated total
volume of Silk Road transactions with the estimated total volume of
transactions at all Bitcoin exchanges (including Mt.Gox~\cite{mtgox},
but not limited to it). The latter corresponds to the amount of money
entering and leaving the Bitcoin network, and statistics for it are
readily available \cite{bitcoincharts}.

Over our entire measurement interval, using the 29-day moving averages
earlier described, approximately 1,335,580 BTC were
exchanged on Silk Road. Likewise, computing 29-day moving averages
for all transactions recorded in Bitcoin exchanges, 
approximately 29,553,384 BTC were traded in Bitcoin exchanges over the
same period. This number is larger than the entire supply of
currency, since each coin is exchanged several times. Comparing
the two numbers shows that Silk Road transactions correspond to about
4.5\% of all transactions occurring in exchanges. 

This estimate is however not particularly robust. Indeed, for each
item purchased on Silk Road, the buyer has to procure, directly or
indirectly, Bitcoins from an exchange,\footnote{Buyers
could have also mined Bitcoins, but we hypothesize this is comparatively
rare.} and eventually the seller may want to redeem funds in their
currency, which corresponds to two exchange transactions. Alternatively, 
sellers may directly re-invest Bitcoins, e.g., to purchase other
items on Silk Road, in which case at most one exchange transaction
occurs.

The only conclusion we can draw from this comparison is that Silk
Road-related trades could plausibly correspond to 4.5\% to 9\% of
all exchange trades. While far from being a negligible share of the
Bitcoin economy, speculative trades (i.e., using
Bitcoin as a commodity rather than a currency) still constitute the bulk
of all exchange trades. A much more thorough analysis would be required
to properly assess the various components of the Bitcoin economy, but is
outside the scope of this paper.

\section{Discussion}
\label{sec:discuss}
While we believe this research may bring up a number of discussions,
and hopefully even start a public policy debate on the effectiveness of
current intervention or prevention strategies for controlled substance
abuse, we here choose to narrow our focus to three areas. First, we discuss the accuracy of our estimates. Second we
provide an overview of the ethical considerations and associated
conclusions we came to during the design of this study. Third, we
briefly evaluate potential intervention policies.

\subsection{Measurement accuracy of sales volumes}
Sales volumes described in Section~\ref{sec:econ} are obtained
through indirect indicators (buyer feedback), and as such only represent
an estimate, and not actual, verified volumes. 

First, dishonest sellers
could plausibly create a number of buyer accounts, leave fabricated
feedback to enhance their reputation, and cause our estimates to be
inflated. Such behavior, however, is usually relatively easy to spot,
and would be reported in Silk Road user forums dedicated to reviewing
sellers. From casual observation, 
the risk of ruining one's reputation for little gain appears
to be enough of a deterrent that fabricated feedback
only represents a small fraction of all feedback and does not impact our
measurements.

On the other hand, buyers typically leave only one piece of feedback
{\em per order}. When an order contains a large quantity of a given
item, we are underestimating the total volume of sales taking place.
Because this cannot be detected, feedback reports only provide a
conservative estimate of overall sales volume. We also cannot account
for sales coming from stealth listings. Thus, it is likely that the
numbers presented in Section~\ref{sec:econ} are a lower bound.

\subsection{Ethical considerations}
Conducting this research yielded some ethical quandaries. Since we are
analyzing data from activity that is, in most jurisdictions, deemed as
criminal, could this work directly lead to arrest or prosecution of
individuals? If such were the case, should it be published?

We answered negatively to the first question. Indeed, the data we
collected is essentially public. We did have to create an account on
Silk Road to access it; but registration is open to anybody who connects
to the site. We did not compromise the site in any way. Perhaps,
bypassing the authentication mechanism and associated CAPTCHA by reusing
an authentication cookie could be construed as a ``hack.'' However, we
argue this is nothing more than using a convenient feature that the site
operators have willingly offered their visitors. Indeed, nothing would
prevent the site operators from setting authentication cookies with very
short expiry dates.

Considering that the data we obtained is available to anybody, we do not
think this work adds any additional risk for the Silk Road operators,
their customers or their sellers. In fact, as routinely expressed in
user forums \cite{silkroadforums}, Silk Road operators seem to espouse
crypto-anarchist ideals (similar to, for instance, 
those described by May \cite{May1994}) and to that end,
willingly make their website -- and, as a result, its data -- publicly
available.

\bulletpoint{Research reproducibility.} At the same time, we need to allow others to reproduce our
results. To that effect, we make a subset of the databases we constructed from our data
collection available at \website. We decided not to make available any
textual information (item name, description, or feedback text) because
we could not manually inspect each entry to ensure that no potentially
private information (e.g., URLs, email addresses) would be released.
Among the results presented in this paper, only the numbers on early
finalization or stealth listings require textual fields. All other
results should be reproducible with the publicly-released dataset.

\bulletpoint{Using the Tor network for measurement research}
Another ethical consideration is linked to the design of the study
itself. To acquire the measurements we needed to obtain for analysis, we
had to repeatedly crawl the Silk Road website. This in turn resulted in
extensively using the Tor network for the purpose of this work. Because
Tor is a relatively resource-constrained network, our measurements
could have impacted other users that would need it. We believe that the
scientific value of this study, and its potential public policy impact
justified the use of the network we have made. Monitoring our usage, we
realized that we were downloading between 500 and 800 MB of data per day
over the Tor network for this project. While certainly not negligible,
this remains in the order of a single typical movie download. 

However, we do think our usage of the network (for this project, and
related research efforts) should be compensated. Partly for this reason,
we have recently deployed a fast Tor relay in our institution's network.

\subsection{Potential intervention strategies}
Given the nature of the goods sold on Silk Road, it is quite clear that
various law enforcement agencies have a strong interest in 
disrupting Silk Road operations. They appear, so far, to have been
unsuccessful since the site is still up and has grown in size since Sen.
Schumer called on the U.S. Attorney General and the head of the U.S.
Drug Enforcement Agency to put an end to it.

We discuss four possible intervention strategies that could be
considered: disrupting the network, disrupting the financial
infrastructure, disrupting the delivery model, and laissez-faire.

\bulletpoint{Attacking the network}
The first possible intervention policy is to disrupt the Tor network.
Indeed, without Tor, Silk Road cannot operate. This strategy is
very likely to be difficult to put in place. First, Tor has many
uses beneficial to society -- Silk Road and other anonymous online
marketplaces are far from representing the majority of Tor traffic,
even though this work argues that their importance is growing. Tor is
routinely used by oppressed individuals to communicate without fear of
reprisal. Thus, disrupting the entire Tor network for the purpose of
taking down Silk Road would come at a high collateral cost.

Furthermore, Tor has shown to be resilient to a large number
of attacks, due to its open design and to the large amount of
academic research it fosters. In particular, Tor hidden services,
like Silk Road, have been the subject of considerable scrutiny
\cite{Overlier:S&P06,Murdoch:CCS06}. {\O}verlier and Syverson showed
that timing and intersection attacks could be used to reveal the
location of hidden services. Most of these concerns have been addressed
in recent versions of Tor, e.g., through the use of persistent ``entry
guards.'' Murdoch described how covert channels (specifically, clock
skew) could leak information allowing to roughly estimate the location of a
hidden service. 

While a determined adversary, given enough measurements, may be able
to roughly estimate the location of a hidden service,
pinpointing its exact location, and then being able to prove that the
machine is in fact hosting the hidden content is considerably more 
challenging if, as we expect, operators of the hidden service suspect
they are being monitored \cite{Shebaro:DFRWS10}. For instance, the
hidden service could merely act as a proxy to another machine hosted
somewhere else.

\bulletpoint{Attacking the financial infrastructure}
Another possible disruption strategy is to attack the financial
infrastructure supporting Silk Road. Bitcoin has shown, in the past,
to be a very volatile currency. The June 2011 theft of a large number
of Bitcoins from the Mt.Gox exchange \cite{mtgox} actually caused an
abrupt collapse of the currency. Certain users have been complaining
in forums of the uncertainty on the prices they end up paying due to
the instability of Bitcoin and the various commissions they have to
pay to purchase Bitcoins, and then to purchase items on Silk Road
\cite{silkroadforums}.

Thus, an adversary could also attempt to manipulate the currency to create rapid
fluctuations and impede transactions. Besides the obvious collateral
costs associated with such strategies, Silk Road does provide hedging
mechanisms against short-term fluctuations of Bitcoin. These mechanisms
have proven to be enough to allow Silk Road to prosper despite Bitcoin's
high volatility -- but it is unclear how they would fare in the face of a
determined attacker with large monetary resources.

Recent research \cite{arXiv:1107.4524v2} has also shown that
Bitcoin transactions are partially vulnerable to traffic analysis.
Indeed, the history of all transactions is publicly available and
network analysis can help map sets of public keys to individual
users and transactions. Since currency exchanges like Mt.Gox where users
redeem Bitcoins for cash bind public keys to actual identities, Bitcoin
anonymity guarantees are weaker than most Silk Road users seem to
assume, even though additional intermediaries (tumblers) are in place.
In particular, large Silk Road sellers withdrawing massive amounts of
Bitcoins at once may be relatively easily identified, unless they take
additional precautions to hide their tracks.

\bulletpoint{Attacking the delivery model}
Another possible angle of action is to attack the delivery model.
That is, to reinforce controls at the post office and/or at customs
to prevent illicit items from being delivered to their destination.
One interesting finding from is that a large number of sellers seem not to 
worry about seizures:
Most items are marked as shipping
internationally, which means that the risk of package loss or destruction
is viewed as minimal by the sellers. This is certainly an area that
warrants further investigation. In the United States, coordination
between agencies is paramount: Customs (which can inspect mail) need to
work in concert with Drug Enforcement Agency (DEA) and/or Food and Drugs
Administration (FDA), depending on the type of item concerned. Very
often, seized packages are simply destroyed, or returned to the
sender.

\bulletpoint{Laissez-faire}
Finally, a last possible intervention strategy is actually not to
intervene. Politically, this is a questionable proposition, as it may
sound as an admission of weakness. There are however studies that show
that drug abuse prevention is considerably more cost-efficient than
enforcing drug prohibition \cite{Caulkins:RAND}. From a public support
standpoint, recent laws in Colorado and Washington state allow the
use of marijuana in these states. Despite the current incompatibility
between these laws and federal regulations, they are certainly a sign of
an evolution in the public attitude toward drug policies. From an
economics standpoint, the relatively rapidly expanding business
of online anonymous markets such as Silk Road, and the logistic
difficulties in shutting down such markets may further tilt the balance
toward prevention and cure.

As a result, laissez-faire, however untenable it might currently appear
from a policy standpoint, might become more attractive in light of
budget constraints. Although there is no public statement about it,
this could be the strategy currently adopted by law enforcement, seeing
that the marketplace has not met any significant disruption to its
operations, other than transient technical issues, in the nine months we
studied it, while at the same time sales volumes have nearly doubled.

\section{Related work}
\label{sec:related}
From a technical standpoint, this work is closely related to rapidly
growing literature on measuring cybercrime (e.g., \cite{Franklin:CCS07,
Levchenko:S&P11, Motoyama:USENIX11, Motoyama:IMC11, McCoy:PETS08,
McCoy:USENIX12, Moore:FC10, Moore:APWG07, Wang:WWW07, LMC:USENIX11}).
The techniques used in this paper (periodic crawls, use of anonymous
networks) to collect measurements indeed are relatively common to most
work in this field. The main difference between this work and
the related cybercrime literature is the object of the measurements.
Instead of trying to characterize a security attack or the behavior
of an attacker, we are instead trying to describe as precisely as
possible an online marketplace. In that respect, our work shares some
similarities with works that have tried to model transactions on eBay
\cite{Houser:JEMS06,Resnick:AAM2002} or Amazon \cite{Mudambi:MIS10}.
However, we do not focus much on customer reviews to assess seller
reputation but instead primarily use feedback as proxy for sales volume.

At first glance quite similar to our work, McCoy et al. provided a
characterization of traffic using the Tor network by monitoring a Tor
exit node \cite{McCoy:PETS08}. Different from this research, we do not
actually monitor Tor traffic and instead analyze data posted to an
online marketplace. Motoyama et al. \cite{Motoyama:USENIX11} performed
related measurements to evidence the existence of online ``mule''
recruitment schemes in crowdsourcing marketplaces.

A more recent paper \cite{McCoy:USENIX12} uses leaked transaction
databases to precisely estimate the revenues and profits of three
major illicit online pharmacy affiliate networks. Contrary to Silk
Road, these networks heavily resort to spam \cite{Levchenko:S&P11} and
search-engine manipulation \cite{LMC:USENIX11} for advertising. McCoy
et al. find that, between 2007 and 2011 the gross revenues of each of
these illicit affiliate networks range from USD~12.8 million/year to
USD~67.7 million/year \cite{McCoy:USENIX12}. Without any advertising other than word-of-mouth,
Silk Road, with its USD~15 million/year transaction volume, appears
to be comparable in size with these illicit online pharmacy networks.
At the same time, Silk Road caters to a priori more technically
sophisticated users, and proposes an inventory that far exceeds
prescription drugs. As such, it seems to occupy a market niche different
from that of traditional illicit online pharmacies. Interestingly,
online pharmacies have recently seen their payment systems being
targeted \cite{McCoy:CCS12}. Whether or not this will lead
them to move to alternative payment instruments such as Bitcoin, or
will redirect their potential customers to marketplaces like Silk Road
remains to be seen.

Our work is also close in spirit to a number of studies in the drug
policy realm. We particularly single out the work by Caulkins and
Reuter \cite{Caulkins1998} who perform an econometric analysis of (offline) 
drug markets to discuss their structure. Finally, Molnar et al.
\cite{Molnar:NSPW10} noted that a number of techniques used to perform
analysis of controlled substance markets could be applied to online
crime as well. Our paper shows that there may be a convergence between
the two fields. 

\section{Conclusion}
\label{sec:concl}
We have performed what we believe to be the first comprehensive measurement
analysis of one of the largest anonymous online marketplaces, Silk
Road. We performed pilot crawls, and subsequently collected daily
measurements for six months (\startdate--\enddate). We analyzed over
24,000 items, and parsed over 180,000 feedback messages. 
We made anonymized versions of our datasets available 
in a companion website (\website). We were able to
determine that Silk Road indeed mostly caters drugs (although
other items are also available), that it consists of a relatively
international community, and that a large fraction of all items do not
remain available on the site for very long. We further discovered that
the number of active sellers 
and sales volume are increasing, corresponding, when averaged 
over our measurement interval
to slightly over USD~1.2~million/month for the entire marketplace, which
in turn represents around USD~92,000/month in commissions for the Silk
Road operators.  Informed by these measurements, we discussed some of
the possible intervention policies. Using appropriate procedures (e.g.,
persistent entry guards), Tor shows good resilience to de-anonymization
attacks, and, barring operator error, whether it is possible to obtain
conclusive evidence of the exact location of a hidden service such as
Silk Road remains an open question. Economic attacks (e.g., artificially
creating large fluctuations in Bitcoin value), while probably more
effective at impeding commerce on such underground marketplaces, would
present significant collateral costs. Ultimately, reducing consumer
demand (e.g., through prevention campaigns) is probably the most viable
strategy.

\section{Acknowledgments}
\label{sec:ack}
We thank our anonymous reviewers, anonymous posters in Bitcoin and Silk
Road forums, Gonzague Gay-Bouchery, Steven Kadlec, Tyler Moore, 
Runa Sandvik, and Zooko Wilcox-O'Hearn for feedback and discussions
on earlier revisions of this manuscript. This research was partially
supported by CyLab at Carnegie Mellon under grant DAAD19-02-1-0389 from
the Army Research Office, and by the National Science Foundation under
ITR award CCF-0424422 (TRUST).

\end{document}